\documentclass[preprint,12pt]{elsarticle}
\usepackage{amsmath}

\usepackage{amssymb}
\usepackage{color,soul}
\usepackage{tablefootnote}
\usepackage[table]{xcolor}
\usepackage{makecell}
\usepackage{multirow}
\usepackage[showframe=false]{geometry}
\usepackage{changepage}
\usepackage{url}
\usepackage[table]{xcolor}
\usepackage{diagbox}
\colorlet{shadecolor}{gray!30}
\colorlet{shadecolor2}{gray!20}
\usepackage{array}
\newcolumntype{?}{!{\vrule width 1pt}}
\usepackage{xr}
\makeatletter
\newcommand*{\addFileDependency}[1]{
  \typeout{(#1)}
  \@addtofilelist{#1}
  \IfFileExists{#1}{}{\typeout{No file #1.}}
}
\makeatother

\newcommand*{\myexternaldocument}[1]{
    \externaldocument{#1}
    \addFileDependency{#1.tex}
    \addFileDependency{#1.aux}
}
\myexternaldocument{Supplementary_material}
\begin{document}

\begin{frontmatter}



\title{H-SynEx: Using synthetic images and ultra-high resolution\\ \textit{ex vivo} MRI for hypothalamus subregion segmentation}
%\uge{Something with \textit{ex vivo}/ ultra-high resolution in it would be good, me thinks}


\author[1,2]{Livia Rodrigues \corref{corauthor}}
\author[3,4]{Martina Bocchetta}
\author[2]{Oula Puonti}
\author[2]{Douglas Greve}
\author[5]{Ana Carolina Londe}
\author[5]{Marcondes França}
\author[5]{Simone Appenzeller}
\author[2,6,7]{Juan Eugenio Iglesias\corref{contrib}}
\author[1]{Leticia Rittner\corref{contrib}}

\cortext[corauthor]{Corresponding author: l180545@dac.unicamp.br}
\cortext[contrib]{Authors contributed equally}
\affiliation[1]{Universidade Estadual de Campinas, School of Electrical and Computer Engineering}
\affiliation[2]{Massachusetts General Hospital, Harvard Medical School}
\affiliation[3]{Dementia Research Centre, Department of Neurodegenerative Disease, UCL Queen Square Institute of Neurology, University College London, London, United Kingdom}
\affiliation[4]{Centre for Cognitive and Clinical Neuroscience, Division of Psychology, Department of Life Sciences, College of Health, Medicine and Life Sciences, Brunel University London, United Kingdom}
\affiliation[5]{Universidade Estadual de Campinas - School of Medical Sciences}
\affiliation[6]{Centre for Medical Image Computing, University College London}
\affiliation[7]{Computer Science and Artificial Intelligence Laboratory,  Massachusetts Institute of Technology}

% \affiliation[inst1]{organization={Department One},%Department and Organization
%             addressline={Address One}, 
%             city={City One},
%             postcode={00000}, 
%             state={State One},
%             country={Country One}}

% \author[inst2]{Author Two}
% \author[inst1,inst2]{Author Three}

% \affiliation[inst2]{organization={Department Two},%Department and Organization
%             addressline={Address Two}, 
%             city={City Two},
%             postcode={22222}, 
%             state={State Two},
%             country={Country Two}}
%for the Frontotemporal Lobar Degeneration Neuroimaging Initiative*

\begin{abstract}
The hypothalamus is a small structure located in the center of the brain and is involved in significant functions such as sleeping, temperature, and appetite control. Various neurological disorders are also associated with hypothalamic abnormalities.  Automated image analysis of this structure from brain MRI is thus highly desirable to study the hypothalamus in vivo. However, most automated segmentation tools currently available focus exclusively on T1w images. In this study, we introduce H-SynEx, a machine learning method for automated segmentation of hypothalamic subregions that generalizes across different MRI sequences and resolutions without retraining.
H-synEx was trained with synthetic images built from label maps derived from ultra-high resolution \textit{ex vivo} MRI scans, which enables finer-grained manual segmentation when compared with 1$mm$ isometric \textit{in vivo} images. We validated our method using Dice Coefficient (DSC) and Average Hausdorff distance (AVD) across \textit{in vivo} images from six different datasets with six different MRI sequences (T1, T2, proton density, quantitative T1, fractional anisotrophy, and FLAIR). Statistical analysis compared hypothalamic subregion volumes in controls, Alzheimer’s disease (AD), and behavioral variant frontotemporal dementia (bvFTD) subjects using the Area Under the Receiving Operating Characteristic curve (AUROC) and Wilcoxon rank sum test. Our results show that H-SynEx successfully leverages information from ultra-high resolution scans to segment \textit{in vivo} from different MRI sequences. Our automated segmentation was able to discriminate controls versus Alzheimer's Disease patients on FLAIR images with 5$mm$ spacing. H-SynEx is openly available at \url{https://github.com/liviamarodrigues/hsynex}.

\end{abstract}


%\section{Abreviation}
%DSC= Dice Coefficient; AVD = Average Hausdorff Distance; AUROC = Area Under the Receiving Operating Characteristic Curve;  AD = Alzheimer`s Disease, bvFTD = behavioral variant frontotemporal dementia; TIV = total intracranial volume

%\section{Summary}
%We propose H-SynEx, a segmentation method for the hypothalamus and its subregions trained on synthetic images derived from high-resolution \textit{ex vivo} MRI, which is compatible with different sequences and resolutions of \textit{in vivo} MRI.

\begin{highlights}
    \item The development of a fully automated segmentation method trained on synthetic images derived from \textit{ex vivo} MRI label maps capable of identifying hypothalamic subregions across various MRI sequences and resolutions, including clinical acquisitions with large slice spacing;
    
     \item The usage of ultra-high resolution \textit{ex vivo} images to build the label maps yields a highly accurate model of the hypothalamus anatomy. 
     
    \item H-SynEx outperforms other state-of-the-art methods in two patient-control comparisons conducted in this study and is currently the only method capable of segmenting hypothalamic subregions on MRI sequences other than T1w and T2w.
\end{highlights}

\end{frontmatter}

\section{Introduction}

\label{sample1}

The hypothalamus is a small, cone-shaped, gray-matter structure located in the central part of the brain. It is composed of subnuclei containing the cell bodies of multiple neuron subtypes. Despite its small dimensions, the hypothalamus plays a significant role in controlling sleep, body temperature, appetite, and emotions, among other functions~\cite{neudorfer2020high, saper2014hypothalamus}. In the literature, several studies establish a connection between the whole hypothalamus and neurodegenerative diseases such as Alzheimer’s disease~\cite{piyush2014analysis}, Huntington's disease~\cite{gabery2015volumetric, bartlett2019investigating}, Behavioral Variant Frontotemporal Dementia (bvFTD) ~\cite{bocchetta2015detailed, piguet}, Amyotrophic Lateral Sclerosis (ALS)~\cite{gorges2017hypothalamic, ahmed}, among others~\cite{seong2019hypothalamic, modi2019individual, wolfe2015focal, gutierrez1998hypothalamic}. Some studies suggest a differential involvement of the hypothalamic subregions across conditions~\cite{bocchetta2015detailed}, leading to the belief that studying these subregions individually is essential for a better understanding of these conditions.

MRI enables the study of the human brain \textit{in vivo}, but many analyses (e.g., volumetry) require manual segmentations that are challenging and time-consuming. For the hypothalamus, manual segmentation is particularly prone to high inter- and intra-rater variability due to its small size and low contrast with neighboring tissue~\cite{billot2020automated, rodrigues2022benchmark, estrada2023fastsurfer}. Even with the help of semi-automated methods, a segmentation of a single scan can take up to 40 minutes~\cite{wolff2018semi}, making large-scale studies impractical at most research sites. To better understand the role of the hypothalamus, several studies use different MRI sequencies~\cite{gorges2017hypothalamic, seong2019hypothalamic, wolfe2015focal, piyush2014analysis, schur2015radiologic}. However, these studies are limited to select sites and require specialists with neuroanatomical knowledge to perform manual annotation. 

Numerous supervised methods have been proposed for the hypothalamus automated segmentation on T1w~\cite{rodrigues2020hypothalamus, billot2020automated, estrada2023fastsurfer, rodrigues2022benchmark, greve2021deep} and T2w images~\cite{estrada2023fastsurfer}. However, none of these methods can segment images at anisotropic resolution (often the case in clinical MRI) or in different sequences than the ones they were trained on (T1w/T2w). They all require retraining to function across different sequences and resolutions, necessitating more labeled data. The use of semi-supervised models on medical images enhances the generalization of networks without necessarily increasing the quantity of annotated data~\cite{fayjie2022semi, bortsova2019semi}. However, most of these models work only in one type of MRI sequence and usually need retraining to adapt to different sequences. Synthetic images allow the construction of training datasets and flawless ground truths~\cite{thambawita2022singan, billot2023synthseg} and the development of methods capable of generalizing in across different MRI sequences~\cite{iglesias2023easyreg, billot2023synthseg}.

So far, all automated hypothalamus segmentation methods were conducted using manual segmentation of \textit{in vivo} images with resolutions ranging between 0.8$mm$ and 1$mm$. Being a small structure, the delineation of the hypothalamus is significantly affected by partial volume effects, even in high-resolution images (such as 0.8$mm$). Recently, the usage of ultra-high resolution \textit{ex vivo} MRI has proven to be beneficial in the segmentation of small structures such as the hippocampus, amygdala, and thalamus~\cite{iglesias2015computational, saygin2017high, iglesias2018probabilistic}, as it permits a better visualization of their anatomical boundaries, leading to more accurate manual annotation

In this article, we train a model using synthetic images derived from label maps built from ultra-high resolution \textit{ex vivo} MRI. Using synthetic images provides robustness againt changes in MRI contrast, while constructing the label maps from \textit{ex vivo} images provides more accurate delineation of the hypothalamus at higher resolution, enhancing the automated segmentation quality. 

H-SynEx, our automated method for hypothalamic subregion segmentation, demonstrates robustness across different MRI contrasts and resolutions. In our experiments, we evaluate its resilience across T1w, T2w, PD, qT1, FA, and FLAIR sequences, as well as in data with 5$mm$ spacing.

\section{Data}
\subsection{Training Data}
\label{data}
\begin{figure} [!ht]  
\begin{center}  
	\includegraphics[width=0.70\columnwidth]{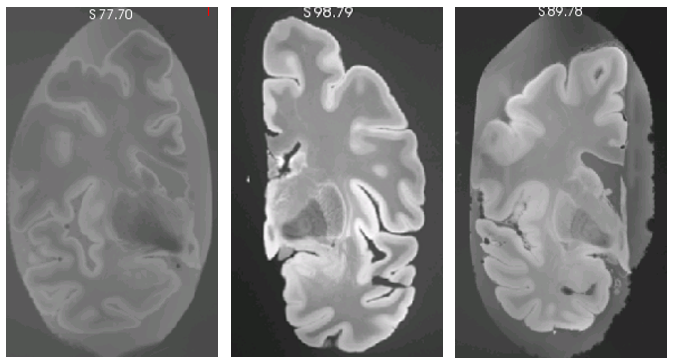}
	\caption{\textbf{\textit{ex vivo} MR images}: Examples of three images used during the method development}
\label{fig:exvivo} 
\end{center}  
\end{figure}

The data used for training H-SynEx comprises synthetic images derived from 3D segmentations (label maps). These label maps are built using a dataset consisting of 10 \textit{post mortem} MRI acquisitions of brain hemispheres~\cite{costantini2023cellular} of 5 male and 5 female specimenss who died of natural causes with no clinical diagnoses or neuropathology.  The voxel resolution ranges from 120 to 150 $\mu$m. The age at the time of death ranges from 54 to 79 years, with an average of 66.4 $\pm$ 8.46 years. The dataset is publicly available at the Distributed Archives for Neurophysiology Data Integration (DANDI Archive)\footnote{\url{https://dandiarchive.org/dandiset/000026/draft/files?location=}}~\cite{dandi} (Figure~\ref{fig:exvivo}).
%We manually labeled five hypothalamic subregions for each of the 10 acquisitions: anterior-superior, anterior-inferior, tuberal-superior, tuberal-inferior, and posterior. We also labeled the fornix, which helps produce more realistic synthetic images. Further details on the manual segmentation, label maps, and synthetic image generation are detailed in Section~\ref{data_preprocessing}

\subsection{Test Data}
\begin{figure} [!ht]  
\begin{center}  
	\includegraphics[width=0.92\columnwidth]{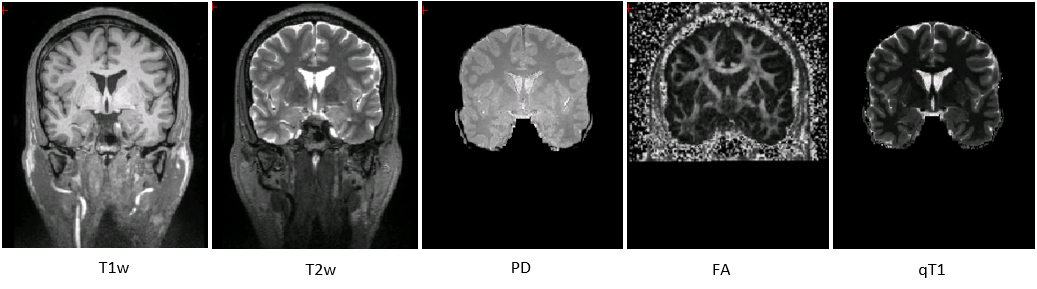}
	\caption{Example of different modalities (FSM dataset) \label{fig:modalities}}   
\end{center}  
\end{figure}

The method evaluation relies on \textit{in vivo} images from 6 different datasets (Table~\ref{tab:datasets}): 

\begin{itemize}
    \item \textbf{FreeSurfer Maintenance (FSM)}~\cite{greve2021deep}: Composed of 29 subjects from which 7 were used for validation and 22 for testing. For each subject, we have T1-weighted (T1w), T2-weighted (T2w), proton density (PD), fracitional anisotropy (FA), and quantitative T1 (qT1) acquisitions (Figure~\ref{fig:modalities}). In all cases the voxel resolution is 1$mm$ isotropic. FSM contains manual labeling for the whole hypothalamus and its subregions (right and left anterior-superior, anterior-inferior, tuberal-superior, tuberal-inferior, and posterior). The manual segmentation was performed on \textit{in vivo} images, and thus with limited accuracy. This dataset was approved by the Massachusetts General Hospital Internal Review Board for the protection of human subjects and all subjects gave written informed consent.  
    
    \item \textbf{MiLI}~\cite{rodrigues2022benchmark}: The MICLab-LNI Initiative comprises manual and automated segmentations of the entire hypothalamus conducted on T1w images with slice thickness between 0.9$mm$ and 1.2$mm$. However, it lacks segmentations for hypothalamic subregions. It includes subjects from various open datasets such as MiLI, OASIS~\cite{lamontagne2019oasis}, and IXI~\cite{IXI}. We only used the manually segmented images, totaling 55 from MiLI (30 controls and 25 ataxia patients), 23 from OASIS, and 19 from IXI. For the latter dataset, as it also encompasses T2w and proton density (PD) acquisitions, we incorporated these modalities in our experiments.  
    
    \item \textbf{ADNI}~\cite{mueller2005alzheimer}: We used a total of 572 controls (280 male and 292 female with average age of $75.5\pm6.4$ and $73.6\pm6.01$, respectively) and 271 Alzheimer's disease (AD) patients (143 male and 98 female with average age of $75.34\pm7.6$ and $73.8\pm7.6$, respectively) for both T1w (1$mm$ isometric) and FLAIR ($0.85mm\times0.85mm\times5mm$) modalities. The ADNI dataset does not have manual segmentation of the hypothalamus.
    
    \item \textbf{NIFD}~\cite{NIFD}: From the Neuroimaging in Frontotemporal Dementia dataset, we used 111 controls (49 male and 62 female with average age of $61.8\pm7.4$ and  $63.4\pm7.8$, respectively)  against 74 behavioral variant frontotemporal dementia (bvFTD) patients (51 male and 23 female with average age of $61.16\pm5.8$ and $62.4\pm7.7$, respectively). The voxel resolution is 1$mm$ isotropic. The NIFD dataset does not have manual segmentation of the hypothalamus.
\end{itemize}

\begin{table}[!ht]
\begin{center}
\caption{\label{tab:datasets} Datasets used for model validation and testing; WS: Whole Structure, SR: Subregion}
\resizebox{1\textwidth}{!}{
\begin{tabular}{|c|c|c|c|c|c|c|c|}
\hline
%{\textbf{}}&{\textbf{Dataset}}&{\textbf{MRI}}&{\textbf{Acquis.}}&{\textbf{Number of}}&{\textbf{Manual }}&{\textbf{Segmentation Protocol}}\\
%{\textbf{}}&{\textbf{Name}}&{\textbf{Sequence}}&{\textbf{Number}}&{\textbf{Subjects}}&{\textbf{Segmentation}}&{WS: Whole Structure}\\
{\textbf{}}&\textbf{\multirow{3}{*}{\shortstack{Dataset \\ Name}}}&
\textbf{\multirow{3}{*}{\shortstack{Sequence \\ type}}}&
\textbf{\multirow{3}{*}{\shortstack{Acquis. \\ Number}}}&
\textbf{\multirow{3}{*}{\shortstack{Subjects \\ Number}}}&
\textbf{\multirow{3}{*}{\shortstack{Voxel \\ Resolution}}}&
\textbf{\multirow{3}{*}{\shortstack{Manual \\ Segmentation \\ Content}}}&
\textbf{\multirow{3}{*}{\shortstack{Segmentation \\ Protocol}}}\\
{}&{}&{}&{}&{}&{}&{}&{}\\
{}&{}&{}&{}&{}&{}&{}&{}\\
\hline

{\multirow{2}{*}{\textbf{Validation}}}&\multirow{2}{*}{FSM}&{T1w, T2w, PD}&\multirow{2}{*}{35}&\multirow{2}{*}{7 Controls}&\multirow{2}{*}{1$mm$ isometric}&\multirow{2}{*}{WS/SR}&{WS:Author~\cite{greve2021deep}} \\
\multirow{10}{*}{\textbf{Testing}}&{}&{FA, qT1}&{}&{}&&{}&{SR:Bocchetta~\textit{et al}~\cite{bocchetta2015detailed}}\\
\Xhline{8\arrayrulewidth}
{}&\multirow{2}{*}{FSM}&{T1w, T2w PD,}&\multirow{2}{*}{110}&\multirow{2}{*}{22 Controls}&\multirow{2}{*}{1$mm$ isometric}&\multirow{2}{*}{WS/SR}&{WS:Author~\cite{greve2021deep}} \\
{}&{}&{FA, qT1}&{}&{}&&{}&{SR:Bocchetta~\textit{et al}~\cite{bocchetta2015detailed}}\\

\cline{2-8}
{}&\multirow{2}{*}{MiLI}&\multirow{2}{*}{T1}&\multirow{2}{*}{55}&{30 Controls}&\multirow{2}{*}{slice thickness between}&\multirow{4}{*}{WS}&\multirow{4}{*}{WS:Rodrigues~\textit{et al}~\cite{rodrigues2022benchmark}}\\
{}&{}&{}&{}&{25 Patients}&&{}&\\

\cline{2-5}
{}&{MiLI-OASIS}&{T1}&{23}&{23 Controls}&{0.9$mm$ and 1.2$mm$}&&\\

\cline{2-5}
{}&{MiLI-IXI}&{T1w, T2w, PD}&{57}&{19 Controls}&{}&&\\

\cline{2-8}
{}&\multirow{2}{*}{ADNI}&\multirow{2}{*}{T1w, FLAIR}&\multirow{2}{*}{1686}&{572 Controls}&{1$mm$ isometric (T1w)}&&\multirow{4}{*}{\textbf{---}}\\
{}&{}&{}&{}&{271 AD Patients}&$0.85\times0.85\times5mm$ (FLAIR)&{No manual}&\\

\cline{2-6}
{}&\multirow{2}{*}{NIFD}&\multirow{2}{*}{T1}&\multirow{2}{*}{185}&{111 Controls}&\multirow{2}{*}{1$mm$ isometric}&{segmentation}&\\
{}&{}&{}&{}&{74 bvFTD patients}&&{}&\\
\hline

\end{tabular} 
}
\end{center}
\end{table}

\section{Methods}
\subsection{Preprocessing  of \textit{ex vivo} MRI}
\label{data_preprocessing}
\begin{figure} [!ht]  
\begin{center}  
	\includegraphics[width=0.85\columnwidth]{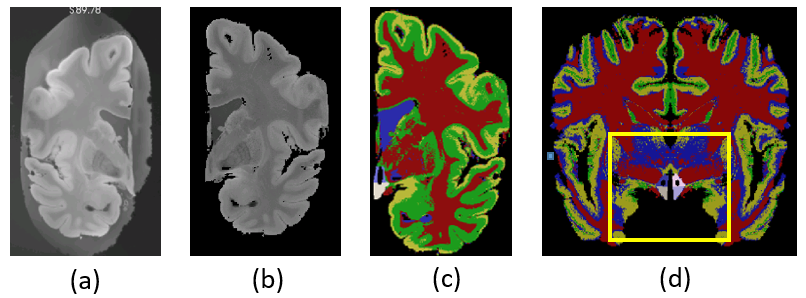}
	\caption{\textbf{Image preprocessing and label maps creation}. (a) Original ~\textit{ex vivo} image. (b) Preprocessed image (c) Automated brain segmentation ($k$=4) and manual hypothalamus segmentation merged (d) The final label map is cropped around the hypothalamus (yellow box) to generate the synthetic images.}
\label{fig:label_maps_creation} 
\end{center}  
\end{figure}

We will train our neural networks with synthetic images generated from label maps. In order to create these, some operations were first performed on the \textit{ex vivo} scans::
\begin{itemize}

\item~\textbf{Preprocessing}: we reoriented the images to conform to positive RAS standards, flipped the right hemispheres, eliminated of all non-brain voxels and performed bias field correction. Also, we resampled the voxels to 0.3 $mm$ to find a balance between high resolution and computational cost. (Figure~\ref{fig:label_maps_creation}(a,b)).

\item~\textbf{Creation of label maps}: Starting with the preprocessed MRI data, we manually delineated the hypothalamus and its subregions. Also, we needed the whole-brain segmentation to bring context around the hypothalamus. However, as other brain structures are not the primary focus of segmentation, it is not necessary for their segmentation to be performed manually, as they may contain noise and may not directly correspond to brain structures. Therefore, we generated automated whole-brain segmentation using k-means, with the value of $k$ varying from 4 to 9, to introduce more variability into the dataset. Lastly, we merged both manual and automated brain hemisphere segmentation (Figure~\ref{fig:label_maps_creation}(c)) and mirrored it to generate a complete whole-brain label map $L\left [D \times H \times W\right ]$ (Figure~\ref{fig:label_maps_creation}(d)). The mirroring process is conducted using an optimization technique that aims to minimize gaps and overlaps. More details on the label maps creation can be found at Rodrigues~\textit{et al}~\cite{rodrigues2024highres}.

\item~\textbf{Find MNI coordinates}: Given that the hypothalamus is a small structure, the use of spatial priors is helpful during training. To achieve this, we integrate MNI coordinates for each input voxel by registering the label maps into the MNI space. Using the label maps $L\left [D \times H \times W\right ]$, we generate a Gaussian image $G \left [D \times H \times W\right ]$ that simulates a T1w MRI. Subsequently, we registered $G$ to the MNI space using NiftiReg~\cite{modat2010lung} and obtain the MNI coordinates $C\left [3 \times D \times H \times W\right ]$ of the registered image. $C$ serves as an additional input channel during training to support the network.

\item~\textbf{Crop}: We crop $L$ and $C$ around the hypothalamus, resulting in two standardized arrays, $L_{\text{crop}}\left [200 \times 200 \times 200\right ]$ and $C_{\text{crop}}\left [3 \times 200 \times 200 \times 200\right ]$, which corresponds to a field of view of $60\times60\times60mm$. 

\item~\textbf{One-hot array}: We convert $L_{\text{crop}}$ into a one-hot array $L_{\text{one}}\left [V \times 200 \times 200 \times 200\right ]$, being $V$ the number of labels presented on $L_{\text{crop}}$. $V$ varies according to the $K$ labels employed on the whole brain segmentation.
\end{itemize}

\subsection{Manual segmentation of the hypothalamus in training data: \textit{ex vivo} images}
\label{manual_seg}

H-SynEx is capable of segmenting 10 subregions of the hypothalamus, being right and left Anterior inferior, Anterior Superior, Tuberal inferior, Tuberal Superior, and Posterior. However, since the \textit{ex vivo} images present only one whole hemisphere of the brain, the manual segmentation was done in only one side of the hypothalamus. The whole structure and subregion segmentation protocol are based on Rodrigues~\textit{et al}~\cite{rodrigues2022benchmark} and Bocchetta~\textit{et al}~\cite{bocchetta2015detailed}, repectively. We also automatically delineate the fornix, using morphological closing. The details on the manual segmentation protocol used to delineate the hypothalamus are described in~\cite{rodrigues2024highres}.

\subsection{Training}

\subsubsection{Synthetic Images Generation}
The synthetic image generation (Figure~\ref{fig:training}(a)) is performed on the fly during training. At each iteration, one of the training label maps, $L$, is randomly selected. $L$ goes throught the preprocessing presented on Section~\ref{data_preprocessing}, which results on the cropped label map ($L_{\text{crop}}$) and MNI coordinates ($C_{\text{crop}}$). Then, we apply aggressive geometric augmentation that encompasses random crop, rotation, and elastic transformation on both $L_{\text{crop}}$ and $C_{\text{crop}}$.,% ending up with $L_{\text{trans}}\left [V \times 160 \times 160 \times 160\right ]$ and $C_{\text{trans}}\left [3 \times 160 \times 160 \times 160\right ]$, respectively. 
Next, we use the generative model proposed by SynthSeg~\cite{billot2023synthseg} based on Gaussian Mixture Models conditioned on the transformed $L_{crop}$, using randomized parameters for contrast and resolution to create the final synthetic image %$S\left [160 \times 160 \times 160\right ]$ (Figure~\ref{fig:simg}). 
%The target $T\left [V \times 160 \times 160 \times 160\right ]$
The transformed $L_{\text{crop}}$ will be the target used to train the network. To assist training, we use an Euclidean distance map ($E$) derived from the target, which has been proven to help locate boundary features during segmentation tasks~\cite{liu2022region}. $E$ is part of the loss function and is only employed during training, not being necessary during inference.
The final input of the network is the concatenation of the synthetic image and the transformed $C_{\text{crop}}$.

\begin{figure} [!ht]  
\begin{center}  
	\includegraphics[width=0.95\columnwidth]{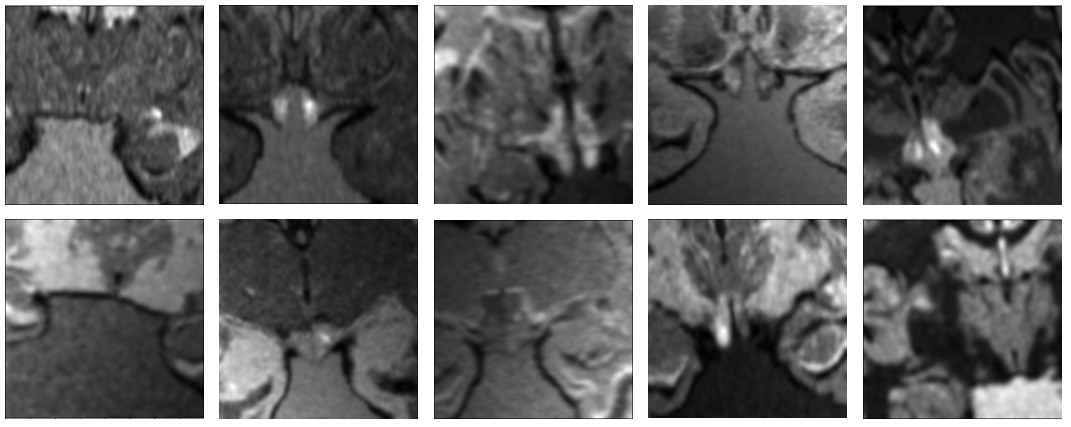}
	\caption{\textbf{Examples of coronal slices from 3D synthetic images used as input}: The images shown here came from the label maps cropped around the hypothalamus. The use of aggressive data augmentation along random contrast values on the generative model results in large variability in the appearance of the input images. \label{fig:simg}}   
\end{center}  
\end{figure}

\subsubsection{Training architecture}
\begin{figure} [!ht]  
\begin{center}  
	\includegraphics[width=0.95\columnwidth]{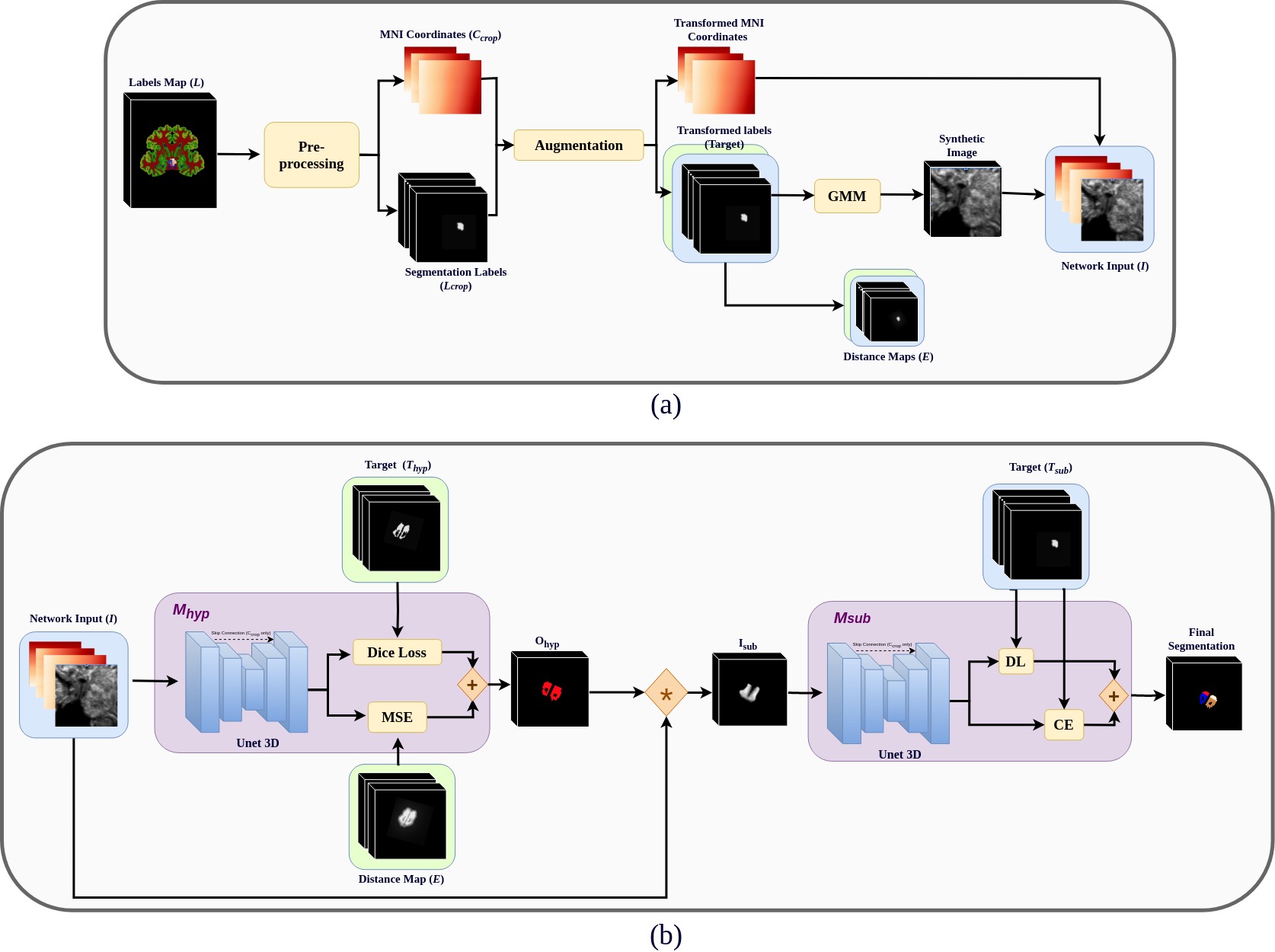}
	\caption{\textbf{Training Flowchart}: (a) Generation of synthetic images: The synthetic images S are generated using the label maps from the \textit{ex vivo} images. (b) Models training: there are two training blocks, one focused on the entire hypothalamus and another specialized in subregion segmentation. The training of the two blocks is done subsequently. We first trained the whole structure segmentation model($M_{hyp}$), and later, the model for the subregions segmentation($M_{sub}$). However, the output of $M_{hyp}$ is used to assist the input creation of $M_{sub}$ during training.  \label{fig:training}}   
\end{center}  
\end{figure}

Two distinct sub-models were trained separately, one for the entire hypothalamus ($M_{\text{hyp}}$) and another specifically for its subregions ($M_{\text{sub}}$) (Figure~\ref{fig:training} (b)). 
Both $M_{\text{hyp}}$ and $M_{\text{sub}}$ are  3D-UNets~\cite{10.7554/eLife.57613, cciccek20163d}, however, in both cases, we added a skip connection between the input channels referring to the transformed $C_{\text{crop}}$ and the final convolutional block to ensure that the original positional encoding is readily available at full-resolution also in the decoder.
$M_{\text{hyp}}$ receives $I$ as input and outputs $O_{\text{hyp}}$. The input of $M_{\text{sub}}$ is defined as $I_{\text{sub}} = I*O_{\text{hyp}}$. 
While $O_{\text{hyp}}$ is a 2-channel array representing the hypothalamus and its background, $O_{\text{sub}}$, the output of  $M_{\text{sub}}$, is a 13-channel array encompassing the subregions, right and left fornices and background.

\subsubsection{Loss function and training details}
The loss function applied to $M_{hyp}$~(\ref{eq:1}) is a combination of Dice Loss ($DL$) and Mean Square Error ($MSE$), while the loss function applied to $M_{sub}$~(\ref{eq:2}), on the other hand, combines $DL$ and Cross Entropy ($CE$). Although our goal is to optimize the Dice coefficient, the Dice loss function has flat gradients away from the optimum at initialization. This issue is mitigated by combining it with other loss functions such as the $MSE$ and the cross-entropy (CE) loss, which provides better gradient information and improves training efficiency. 

\begin{equation}
    \label{eq:1}
    L_{\text{hyp}} = \alpha*DL\left (T, T_{\text{pred}}\right ) + \beta*\text{MSE}\left (E, E_{\text{pred}}\right )
\end{equation}

\begin{equation}
    \label{eq:2}
    L_{\text{sub}} = \alpha*DL\left (T, T_{\text{pred}}\right ) + \beta*\text{CE}\left (T, T_{\text{pred}}\right )
\end{equation}
  
 For both models, we used Adam optimizer with a learning rate of $5*10^{-5}$, a batch size of 32, and values of $\alpha$ and $\beta$ as $0.3$ and $0.7$, respectively. 
As stop criteria, we simply trained $M_{hyp}$ for 40000 training steps and did not use any validation set. However, on $M_{sub}$, we used 35 images from FSM (5 acquisitions from different MRI sequences from 7 distinct subjects) as validation set(Table~\ref{tab:datasets}). We set an early stop criteria based on the DSC of the validation set. For this, we defined the stopping criteria as $\delta_{min}$ = 0.001. The network trained for 28000 steps and stopped.
Both 3D U-Net modules are composed of an encoder of 5 levels with 24, 48, 96, 192, and 384 feature maps. Each convolutional block is composed of three layers: group normalization, convolution, and activation function (ReLU).

\subsection{Inference and Post processing}
\begin{figure} [!ht]  
\begin{center}  
	\includegraphics[width=0.95\columnwidth]{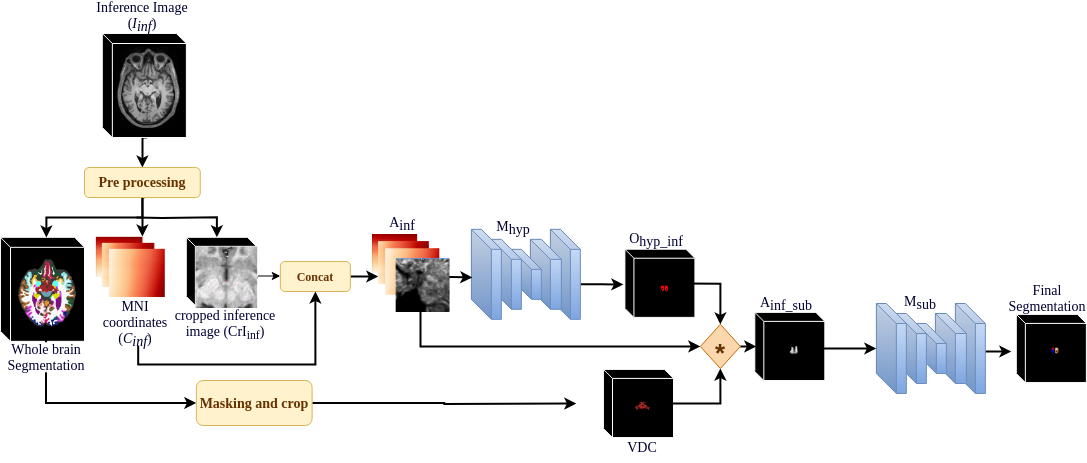}
	\caption{\textbf{Inference flowchart}: The inference image $I_{\text{inf}}$ goes through a preprocessing step to find the input array $A_{\text{inf}}$. $A_{\text{inf}}$ is then applied to the whole structure segmentation model($M_{hyp}$). Finally, using VDC, $A_{\text{inf}}$ and the output of $M_{hyp}$ ($O_{\text{hyp\_inf}}$), we create the input for the subregion segmentation model ($M_{sub}$) and find the final subregions segmentation. 
  \label{fig:inference}}   
\end{center}  
\end{figure}

The inference process is summarized in Figure~\ref{fig:inference}. The first step is preprocessing, in which we find the MNI coordinates ($C_{\text{inf}}$) of the input image, by using a fast deep learning algorithm, EasyReg~\cite{iglesias2023easyreg}. 
The input of $M_{hyp}$, defined as $A_{\text{inf}}$, is found by cropping and concatenating $C_{\text{inf}}$ and the original image ($I_{\text{inf}}$). The input for $M_{\text{sub}}$, however, is formed by the product of $A_{\text{inf}}$, the output of $M_{hyp}$, and the ventral diencephalon (VDC) label, which is derived from the whole brain segmentation produced by EasyReg~\cite{iglesias2023easyreg}.  The inclusion of the ventral-DC label is justified as we found it to reduce false positives within the anterior subregion. The post-processing phase comprises two sequential steps: the rescaling of the final segmentation to match the voxel size of the original image ($I_{\text{inf}}$), and the exclusion of voxels that belong to the third ventricle by using the whole brain segmentation obtained from EasyReg~\cite{iglesias2023easyreg}.

%\subsection{Evaluation Metrics}
%\label{evalMetrics}
%To conduct a quantitative analysis of the results and compare H-SynEx with other state-of-the-art methods, we employed the Dice coefficient (\textit{DSC}) and the average Hausdorff distance (\textit{AVD}). The selection of these metrics was influenced by the specific characteristics of the hypothalamus (a small structure with low contrast) and the established usage of these metrics in the scientific literature.
 
% \begin{itemize}
%     \item \textbf{Dice Coefficient:}

%The \textit{DSC} is an overlap measure defined as follows:
%\begin{equation}
%     DSC= \frac{2*\left | M\cap A \right |}{|M|+ |A|}
% \end{equation}

% \textit{DSC} is sensitive to small structures and does not identify boundary errors. However, it can be used as a measure of reproducibility and is widely used for medical imaging segmentation analysis, being the most used metric in the medical imaging segmentation field.
% \textit{DSC} results range between [0,1] range, where 1 a perfect \textit{DSC}

% \item \textbf{Average Hausdorff Distance}:
    
% \textit{AVD} is the average Hausdorff Distance over all points.

% \begin{equation}
%     AVD(A, M) = max(d(A, M), d(M,A))
% \end{equation}

% where:

% \begin{equation}
%   d(A,M) = \frac{1}{N}\sum_{a\epsilon A}\min_{m\epsilon M}\left \| a-m \right \|
% \end{equation}

% The AVD is a spatial distance metric that unlike \textit{DSC}, is sensitive to boundary errors and more robust than the DSC when analyzing small structures. The smaller the \textit{AVD}, the better the segmentation.

% \end{itemize}

\subsection{Statistical Analysis}
The statistical analysis was done using the AVD and DSC combined with Wilcoxon signed-rank tests to assess the statistical significance of differences in performance across methods. We also compared the ability of H-SynEx and competing methods to find statistical differences in the volume of hypothalamus subregions of controls and patients (AD and bvFTD). Since the datasets have few subjects and we can not assess with high significance that the distribution is Gaussian, the statistical analyses were conducted considering non-parametric distributions. We used Wilcoxon rank-sum test to assess the significant difference in medians between groups and the area under the receiving operating characteristic curve (AUROC) as a non-parametric version of effect sizes between groups. Finally, we used the DeLong test to compare AUROCs across methods operating on the same sample. All statistical tests were conducted with a confidence level of 95\% $(p-value<0.05)$

\section{Experiments and Results}

H-SynEx was trained using synthetic images derived from ultra-high-resolution \textit{ex vivo} label maps. While the synthetic approach increase the network ability to generalize across different sequences, the use of \textit{ex vivo} images improve the ability to delineate the hypothalamus due to their ultra-high-resolution. Given that, our experiments were structured to assess the method's applicability under diverse conditions (Table~\ref{tab:experiments}). 

%Initially, we conducted an inter-rater analysis to establish quantitative metric baselines. Subsequently, we validated the method's performance across different modalities using the FSM dataset (encompassing T1w, T2w, PD, FA, and qT1 sequences) and the IXI dataset (consisting of T1w, T2w, and PD sequences). Following this, we compared H-SynEx with other methods from the literature across multiple datasets (MiLI, MiLI-OASIS, MiLI-IXI, FSM), including one comprising both patient and control subjects (MiLI). We then conducted an application analysis in group studies involving patients with AD and bvFTD (using ADNI and NIFD datasets, respectively). Finally, after confirming the method's usability with images of different sequences, we assessed its performance across various resolutions using FLAIR images from ADNI with a slicing of 5$mm$.
 
\begin{table}[!ht]
\begin{center}
\caption{\label{tab:experiments} Summary of conducted experiments}
\resizebox{1\textwidth}{!}{
\begin{tabular}{|c|c|c|c|c|}
\hline
{}&{}&\multicolumn{3}{c|}{\textbf{Testing set}}\\
\cline{3-5}
{\textbf{\textbf{Experiment}}}&{\textbf{Objective}}&\multirow{2}{*}{\textbf{Dataset}}&\textbf{Number of Acquisitions}&\textbf{MRI}\\
{\textbf{\textbf{}}}&{\textbf{}}&{}&\textbf{per MRI Sequence}&\textbf{Sequence}\\
\hline
{\multirow{2}{*}{\textbf{Inter-Rater Metrics}}}&{To establish a baseline for}&\multirow{2}{*}{FSM}&\multirow{2}{*}{10}&\multirow{2}{*}{T1}\\
{\textbf{}}&{evaluation metrics}&&&\\
\Xhline{5\arrayrulewidth}
{\textbf{Direct comparison with}}&{To assess whereas the method}&\multirow{2}{*}{FSM}&\multirow{2}{*}{22}&{T1w, T2w,PD,}\\
{\textbf{manual segmentation on}}&{is capable to segment }&&&{ FA, qT1}\\
\cline{3-5}
{\textbf{different sequences}}&{on different MRI sequences}&{IXI}&{19}&{T1w,T2w,PD}\\
\Xhline{5\arrayrulewidth}
%MiLI
{}&{Comparing H-SynEx}&{MiLI}&{55}&\multirow{4}{*}{T1}\\
\cline{3-4}
%MiLI-OASIS
{\textbf{Comparing against}}&{against other state-of-the-art}&{MiLI-OASIS}&{23}&{}\\
\cline{3-4}
%MiLI-IXI
{\textbf{state-of-the-art methods}}&{available methods}&{MiLI-IXI}&{19}&{}\\
\cline{3-4}
%FSM
{}&{using only T1 images}&{FSM}&{22}&{}\\
\Xhline{5\arrayrulewidth}
{\textbf{Application in}}&{Assess the method usability}&{ADNI}&{843}&\multirow{2}{*}{T1}\\
\cline{3-4}
{\textbf{Group Studies}}&{in group studies}&{NIFD}&{185}&\\
\Xhline{5\arrayrulewidth}
{\textbf{Resilience to large}}&{To assess usability}&\multirow{2}{*}{ADNI}&\multirow{2}{*}{843}&\multirow{2}{*}{T1w, FLAIR}\\
{\textbf{slice spacing}}&{on diverse MRI resolution}&{}&{}&{}\\

\hline
\end{tabular} 
}
\end{center}
\end{table}

\subsection{Consistency between labeling protocols} 

One of the primary challenges in analyzing the results of our experiment is that each dataset used in testing has a distinct manual segmentation protocol, none of which aligns with the one employed in training H-SynEx due to the difference between \textit{in vivo} and \textit{ex vivo} images~\cite{rodrigues2024highres}. Therefore, our initial experiment aims to establish an upper bound value for DSC and AVD by comparing inter-rater metrics using distinct segmentation protocols performed on T1w images. We compare manual segmentations in 10 FSM images delineated by two different raters: the first uses the FSM protocol (Table~\ref{tab:datasets}) while the second employs the protocol used during the label maps construction(Section~\ref{manual_seg}).

The results (Table~\ref{tab:inter_tab}) shows that despite the AVD is influenced by different protocols, its values remain small, with the highest being 0.43. The DSC metric, however, is affected by both the variations in segmentation protocols and the small size of the hypothalamus subregions, resulting in final values of 0.66 or lower.

\begin{table}[!ht]
\begin{center}
\caption{\label{tab:inter_tab} Inter-rater metrics (median) for 10 subjects from FSM}
\resizebox{0.4\textwidth}{!}{
\begin{tabular}{c|c|c}
{\backslashbox{Subregion \kern-1.5em}{\kern-3.5em Metric}}& {DSC}& {AVD(mm) }\\
\hline
{Anterior}& {0.63}& {0.41}\\
\hline
{Tuberal}& {0.66}& {0.43}\\
\hline
{Posterior}& {0.66}& {0.38}\\
\end{tabular} 
}
\end{center}
\end{table}

\subsection{Direct comparison with manual segmentation on different sequences} 
 
In this experiment, we aim to evaluate the ability of H-SynEx to properly segment the subregions of the hypothalamus in different MRI sequences. We employed five different sequences from FSM - T1w, T2w, proton density (PD), fractional anisotropy (FA), and quantitative T1 (qT1)- and three from IXI -T1w, T2w, and PD. As other methods from the literature exclusively operate on T1w images, a quantitative comparison of their metrics with H-SynEx was not possible in this experiment. 

\begin{figure} 
\begin{center}  
	\includegraphics[width=0.9\columnwidth]{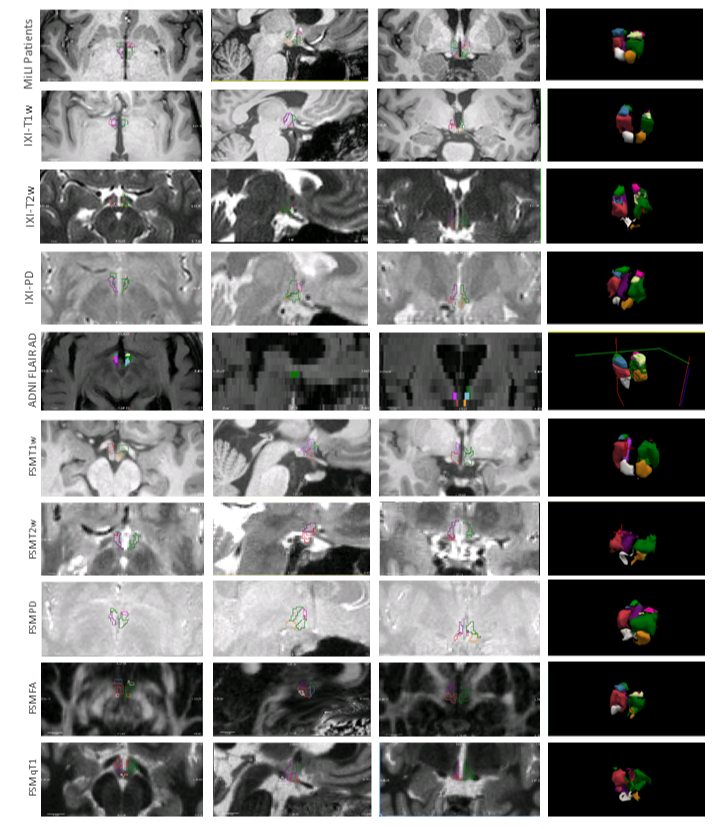}
	\caption{\textbf{Qualitative results} in different datasets, sequences, and resolutions for H-SynEx. Other methods, when applied to sequences different from T1w, return no results \label{fig:qualitative_hidden} }

\end{center}  
\end{figure}

\begin{figure} [!ht]  
\begin{center}  
	\includegraphics[width=1\columnwidth]{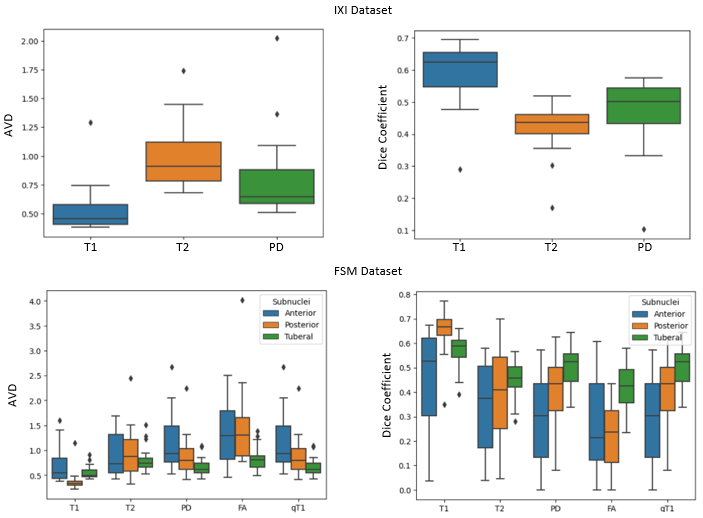}
	\caption{\textbf{DC, and AVD for H-SynEx across diverse sequences}. Top row: IXI dataset, which only presents the segmentation of the whole structure (excluding the mammillary bodies). Bottom row: FSM dataset, that contains the segmentation of the hypothalamus and its subregions. \label{fig:exp1_plots}}   
\end{center}  
\end{figure}

Analyzing H-SynEx metrics on different sequences, we can see that the method presents a better performance on T1w images (Figure~\ref{fig:exp1_plots}). Yet, it is capable of segmenting the hypothalamus and its subregions in all the proposed MRI sequences, as can be seen in Figure~\ref{fig:qualitative_hidden}. 

\begin{figure} [!ht]  
\begin{center}  
	\includegraphics[width=0.6\columnwidth]{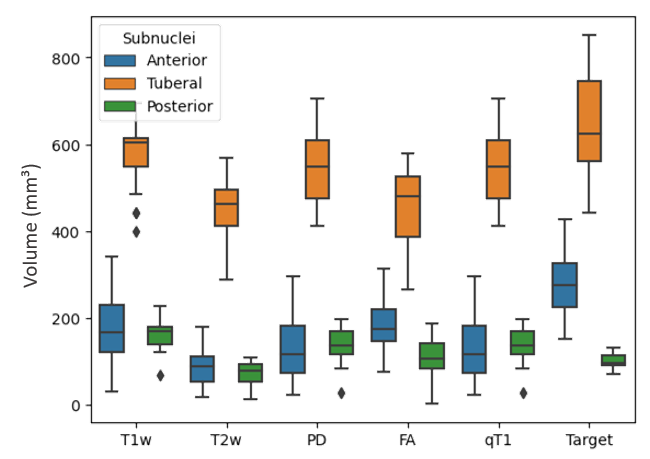}
	\caption{\textbf{Hypothalamus volume}: H-SynEx and manual segmentation (target) volumes for FSM dataset. \label{fig:exp1_vol}}   
\end{center}  
\end{figure}

\subsection{Comparing against other state-of-the-art methods}
To compare H-SynEx with other state-of-the-art models~\cite{billot2020automated,rodrigues2022benchmark, greve2021deep}, we used T1w images from MiLI and FSM datasets and analyzed the whole hypothalamus segmentation. It is worth noting that the MiLI segmentation protocol does not include the mammillary bodies. 
Therefore, for this dataset, we excluded the posterior subregion from the results before computing the metrics. Similarly, HypAST does not segment the posterior subregion, therefore we excluded it from FSM in this case, before running the metrics.

Given that Billot~\textit{et al}~\cite{billot2020automated} works only on T1w images, we compared its results on the hypothalamus suregions with H-SynEx on 22 T1w images from FSM (Table~\ref{tab:metrics_sub}). Finally, to compare H-SynEx with ScLimbic~\cite{greve2021deep} and Rodrigues~\textit{et al}~\cite{rodrigues2022benchmark} we used the whole structure (Table~\ref{tab:metrics}).

\begin{table}[!ht]
\begin{center}
\caption{\label{tab:metrics_sub} AVD and DSC(median) for H-SynEx and Billot~\textit{et al.} on different subregions for FSM dataset. \dag  indicates statistical significance on a two-sided Wilcoxon rank-sum test using Bonferroni correction for $p < 0.05$}.

\resizebox{0.6\textwidth}{!}{
\begin{tabular}{c|c|c|c}
{}&{\backslashbox{Subregion \kern-1em}{\kern-2em Model}}&{H-SynEx}&{Billot~\textit{et al.}}\\
\cline{1-4}
\multirow{3}{*}{\rotatebox[origin=c]{90}{\makecell[c]{\textbf{AVD}\\\textbf{(mm)} \textbf{}}}}&{Anterior}&{\textbf{0.54}\dag}&{1.32}\\
\cline{2-4}
&{Tuberal}&{\textbf{0.49}\dag}&{0.66}\\
\cline{2-4}
&{Posterior}&{\textbf{0.33}\dag}&{0.52}\\
\hline
\hline
\multirow{3}{*}{\rotatebox[origin=c]{90}{\makecell[c]{\textbf{DICE}\\ \textbf{}}}}&{Anterior}&{\textbf{0.53}\dag}&{0.33}\\
\cline{2-4}
&{Tuberal}&{\textbf{0.59}}&{\textbf{0.58}}\\
\cline{2-4}
&{ H-Posterior}&{\textbf{0.67}\dag}&{0.55}\\

\end{tabular} 
}
\end{center}
\end{table}

\begin{table}[!ht]
\begin{center}
\caption{\label{tab:metrics} AVD and DSC(median) for H-SynEx, ScLimbic~\cite{greve2021deep} and Billot~\textit{et al.}~\cite{billot2020automated} on different datasets (MiLI, IXI, OASIS, and FSM) for the entire hypothalamus (except MB). The symbols indicate statistical significance on a two-sided Wilcoxon rank-sum test using Bonferroni correction for $p<0.05$: (*) Billot \textit{vs} H-SynEx; ($^\dag$) ScLimbic \textit{vs} H-SynEx; ($^\ddag$) Billot \textit{vs} ScLimbic. Since ScLimbic was trained using the FSM dataset, we did not consider these results. Similarly, since HypAST was trained using data from MiLI, IXI and the same segmentation protocol as OASIS, we did not consider these results}.
\resizebox{0.6\textwidth}{!}{
\begin{tabular}{c|c|c|c|c|c}
{}&{\backslashbox{Model \kern-1em}{\kern-2em Dataset}}&{MiLI}&{IXI}&{OASIS}&{FSM}\\
\cline{1-6}
\multirow{4}{*}{\rotatebox[origin=c]{90}{\makecell[c]{\textbf{AVD}\\\textbf{(mm)}  \textbf{}}}}&{Billot}&{0.46}&{0.61*$^\ddag$}&{\textbf{0.47}}&{\textbf{0.40}}\\
\cline{2-6}
&{HypAST}&{-}&{-}&{-}&{\textbf{0.41}}\\
\cline{2-6}
&{ScLimbic}&{\textbf{0.39}$^\dag$$^\ddag$}&{\textbf{0.44}}&{\textbf{0.49}}&{\textbf{-}}\\
\cline{2-6}
&{H-SynEx}&{0.45}&{\textbf{0.45}}&{\textbf{0.5}}&{\textbf{0.43}}\\
\hline
\hline
\multirow{3}{*}{\rotatebox[origin=c]{90}{\makecell[c]{\textbf{DICE}\\ \textbf{}}}}&{Billot}&{0.66*}&{0.6}&{\textbf{0.65*$^\ddag$}}&{\textbf{0.68}}\\
\cline{2-6}
&{HypAST}&{-}&{-}&{-}&{\textbf{0.69}}\\
\cline{2-6}
&{ScLimbic}&{\textbf{0.67}$^\dag$$^\ddag$}&{\textbf{0.64}$^\dag$$^\ddag$}&{0.59}&{-}\\
\cline{2-6}
&{ H-SynEx}&{0.63}&{0.62}&{0.58}&{\textbf{0.65}}\\

\end{tabular} 
}
\end{center}
\end{table}

\subsection{Application to group studies}

%. Please say that this setup is representative of what people will be using this tool for.
%2. Also, say that you can use the ability to separate groups as a proxy for performance on datasets that have no ground truth segmentation

In this experiment, we employ H-SynEx on images acquired from both patient and control groups to simulate the real-world application of this method by physicians. Also, we assess the ability of the network to separate groups as a proxy for performance on datasets that have no ground truth segmentation

In the literature, we can find some studies that point to hypothalamic atrophy in both AD and bvFTD patients~\cite{bocchetta2015detailed, tao2021hypothalamic}. Therefore, to evaluate the group studies, we compared the hypothalamic subregion volumes of patients and control groups from ADNI (AD subjects) and NIFD (bvFTD subjects). We normalized the volumes by dividing them by the total intracranial volume (TIV), provided by SynthSeg~\cite{billot2023synthseg}. This normalization is a common practice in volumetric studies with brain MRI. For comparative purposes, we conducted the analysis using Billot~\textit{et al}. and compared with H-SynEx through DeLong test~\cite{delong1988comparing}. 

Observing the applicability of the methods on group studies (Table~\ref{tab:exp3_4}), H-SynEx achieved statistical significance $(p<0.05)$ in the Wilcoxon rank-sum test in all hypothalamic subregions when comparing AD vs. controls, while Billot \textit{et al.} was unable to detect differences in the tuberal-inferior region. Additionally, in some cases, we observed a higher AUROC in H-SynEx, along with a $p-value<0.05$ for the DeLong test, indicating the ability of H-SynEx to better discern differences between the two groups in this dataset. Regarding NIFD, the results were similar for both models, except for the tuberal-inferior region. 

\subsection{Resilience to large slice spacing}
In this experiment, we applied H-SynEx on FLAIR images from the ADNI dataset acquired with a slice spacing (and thickness) of $5mm$ in the axial plane. Here, we want to evaluate our method's capability to identify hypothalamic atrophy with larger spacings, which are common in clinical MRI. Once no other method in the literature works with FLAIR images, we solely compared H-SynEx segmentations on 5$mm$ spacing FLAIR images from the same subjects from the ADNI dataset used in Experiment 4.4.  
When analyzing the volumes, H-SynEx returns statistically significant results (Table~\ref{tab:exp3_4}) when comparing patient and control volumes normalized by TIV in all subregions, except for the posterior subregion.

\begin{figure} [!ht]  
\begin{center}  
	\includegraphics[width=0.9\columnwidth]{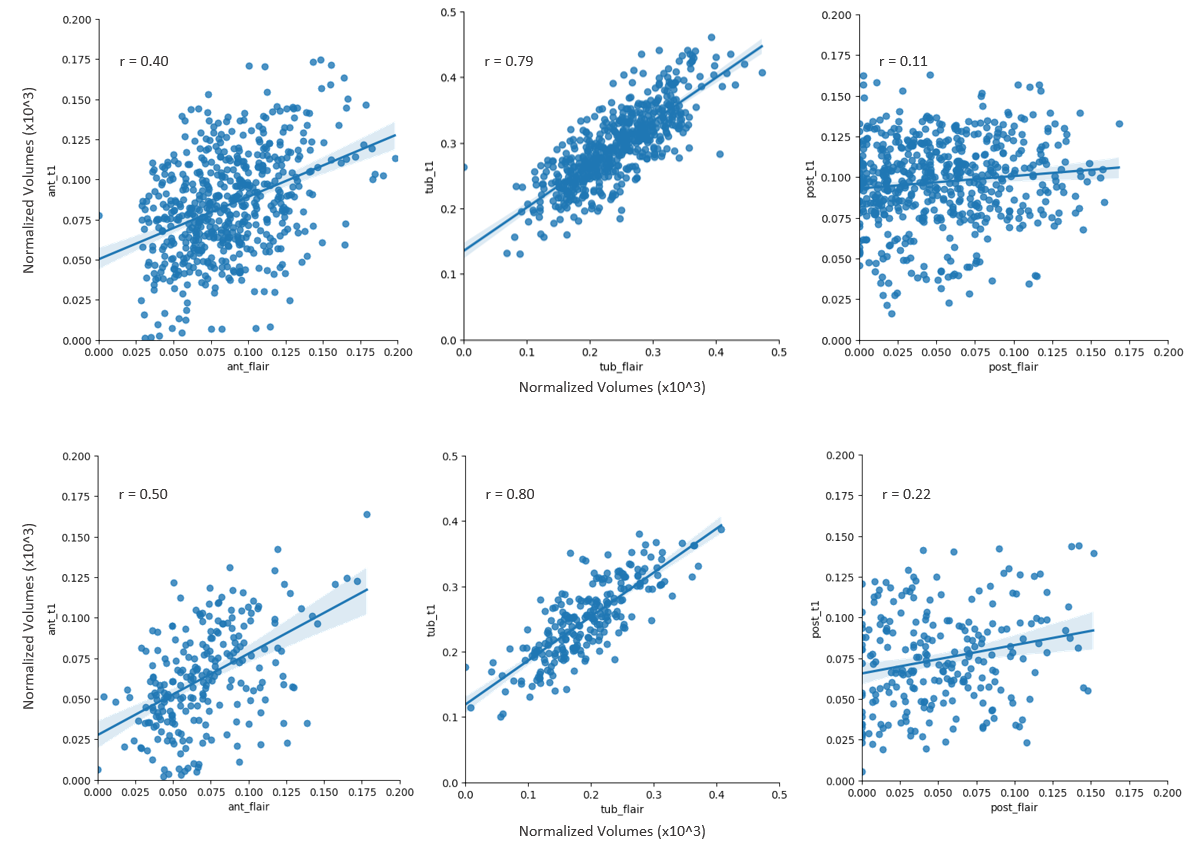}
	\caption{\textbf{Normalized volume correlation for FLAIRs \textit{vs} T1w (ADNI Dataset) using H-SynEx segmentation}. Up: Control subjects; Down: AD patients. We can see that besides the posterior subregion, we can find a positive correlation between FLAIR and T1w normalized volumes.}
\label{fig:exp4_corr} 
\end{center}  
\end{figure}

\begin{table}[!ht]
\setlength{\tabcolsep}{12pt}
\begin{center}
\caption{ AUROC Values for patients vs. controls for H-SynEx and Billot methods in ADNI and NIFD datasets. For ADNI dataset, we also analyze our method when applied to FLAIR images with spacing of $5mm$. Stars indicate the level of statistical significance (two-sided Wilcoxon rank-sum test) between both cohorts (* $p < 0.05$, ** $p < 0.01$). $^\dag$ indicates statistical significance on the DeLong test ($p<0.05$) between H-SynEx and Billot methods. $^\ddag$ indicates statistical significance on the DeLong test ($p<0.05$) between H-SynEx applied on T1-w and H-SynEx applied on Flairs. \label{tab:exp3_4}}
\resizebox{1\textwidth}{!}{
\begin{tabular}{c||c|c|c||c|c}
\textbf{Dataset}&\multicolumn{3}{ c||}{\textbf{ADNI}}&\multicolumn{2}{ c}{\textbf{NIFD}} \\
\hline
\textbf{\backslashbox{Subregion \kern-1.5em}{\kern-3.5em Model}}& \textbf{\makecell{H-SynEx\\Flair}}&\textbf{\makecell{H-SynEx\\T1w}}&\textbf{\makecell{Billot\\T1w}}&\textbf{\makecell{H-SynEx\\T1w}}&\textbf{\makecell{Billot\\T1w}}\\
\hline
{\textbf{Whole}}& {0.66**$^\ddag$ } & {0.74**}& {0.65**$^\dag$}& {0.79**}& {0.74**}\\
\hline
{\textbf{a-sHyp}}& {0.60**$^\ddag$ } & {0.69**}& {0.72**}& {0.76**}& {0.75**}\\
\hline
{\textbf{a-iHyp}}& {0.60**} & {0.64**}& {0.55*$^\dag$}& {0.72**}& {0.62**}\\
\hline
{\textbf{supTub}}& {0.68**$^\ddag$ } & {0.60**}& {0.67**$^\dag$}& {0.76**}& {0.76**}\\
\hline
{\textbf{infTub}}& {0.67**$^\ddag$ } & {0.73**}& {0.52$^\dag$}& {0.74**}& {0.59*$^\dag$}\\
\hline
{\textbf{postHyp}}& {0.52$^\ddag$ } & {0.72**}& {0.70**}& {0.7**}& {0.73**}\\
\end{tabular} 
}
\end{center}
\end{table}

\section{Discussion and Conclusion}
%\livia{Limit to 800 words, currently with 801}
%\uge{please break into paragraphs}
Due to the small size of the hypothalamus and its low contrast compared to neighboring tissues, its manual segmentation is challenging, and variable among and within raters. These characteristics extend across various MRI sequences. To address this issue, we introduced H-SynEx, a novel automated segmentation method for the hypothalamus and its subregions. To the best of our knowledge, H-SynEx is the first method to combine ultra-high-resolution \textit{ex vivo} MRI and synthetic images. This integration has allowed us to develop a method capable of effectively segmenting small structures, such as hypothalamus subregions, across various MRI sequences and resolutions, including FLAIR images with a spacing of 5$mm$.

Typically, when evaluating how well a developed segmentation method generalizes, we compare it to others found in existing literature. To do this, it is common to use a dataset that none of the methods have seen during training. However, when these methods use training sets with different segmentation protocols, this difference can introduce bias, favoring the method trained under the same protocol as the test images.  By using \textit{ex vivo} images to construct the training set, the segmentation protocol used in training H-SynEx became different from any other \textit{in vivo} image set. Consequently, the main challenge in analyzing the results lies in the difference between the training and test protocols. Focusing on that, we compared the manual segmentation of two raters who employed distinct protocols on 10 T1w images from the FSM dataset and found inter-rater DSC values lower or equal to 0.66 and AVD higher or equal to 0.38. We use these values as a baseline for analyzing the metrics in the subsequent experiments.

On Experiment 2, we analyzed H-SynEx usability across different MRI sequences. We could assess that T1w images presented the best results. However, despite the lower DSC and higher AVD values for the other sequences, it is important to emphasize that the manual segmentations of the hypothalamus subregions in both FSM and IXI were done in T1w images, not being influenced by the different contrasts of other sequences. Additionally, while the FSM images for each subject are already registered, this is not the case for the IXI dataset. Hence, the manual segmentations were registered to be used on the different sequences acquisitions of the same subject. Both registration and the use of a different sequence for manual segmentation may compromise the final results. Finally, we could notice a high variability on both metrics, which may be explained by the small size of the hypothalamus. This hypothesis is reassured by comparing the volumes delineated by H-SynEx and manual segmentation in the FSM dataset (Figure~\ref{fig:exp1_vol}). We can see that both the posterior and anterior subregions, which show greater variability in the DSC and AVD, are relatively smaller than the tuberal subregion. Furthermore, the variability in volumes across sequences and subregions appears to be less pronounced than the variability in the metrics. For instance, for the anterior subregions we can see a large variability in the DSC, which is less pronounced in both AVD and the volumetric analysis. This may imply that the small size of the anterior subregion may be interfering in the final DSC values. The same analysis is valid for the posterior region.

When comparing H-SynEx with other state-of-the-art methods, we see that H-SynEx outperforms Billot~\textit{et al} in almost every metric for subregion segmentation. Here, it is important to highlight that despite DSC values seem to be low at first glance, they are not far from the values observed in the inter-rater analysis. H-SynEx AVD values, however, demonstrate greater similarity to inter-rater AVD, particularly in the posterior subregion where even lower AVD values are observed. Additionally, H-SynEx AVD metrics are substantially lower compared to those reported by Billot~\textit{et al}. Observing AVD and DSC for the whole structure (Table~\ref{tab:metrics}), H-SynEx outperforms Billot~\textit{et al} and returns similar results to HypAST~\cite{rodrigues2022benchmark} and ScLimbic~\cite{greve2021deep} on the former, despite not achieving the best performance on the latter. However, when dealing with small structures with complex boundaries, distance metrics such as AVD, are more suitable to compare different methods ~\cite{taha2015metrics}. Also, it is important to emphasize that all other methods were exclusively trained on \textit{in vivo} T1w images, not having to deal with domain gap. Despite not achieving the highest quantitative results on T1w images, H-SynEx offers a distinct advantage. Built upon well-established domain randomization methods, it demonstrates superior generalization ability across MRI sequences. This enhanced robustness stems from its ability to handle variations in data, making it more adaptable to different imaging conditions.

When comparing volumes of the hypothalamus from patient and control groups on T1w images, we have confirmed that our method detects expected differences in all subregions in ADNI and NIFD datasets, with AUROCs of 0.74 and 0.79 respectively, and $p-value<0.05$ for the Wilcoxon signed-rank test in both cases. Notably, the AUROC values reported to NIFD are higher than those found in ADNI (Table~\ref{tab:exp3_4}). This behavior is expected since bvFTD patients tend to exhibit more pronounced hypothalamic atrophy than AD patients (10-12\% volume loss in AD and 15-20\% in bvFTD)~\cite{vercruysse2018hypothalamic}. Additionally, we determined that H-SynEx results differ statistically from Billot~\textit{et al} for the entire hypothalamus and in most subregions in the ADNI dataset, with a $p-value<0.05$ for the DeLong test. 

Finally, we analyzed the same subjects from ADNI used in experiment 4, but using FLAIR images with a spacing of  5$mm$. It is possible to see that, similarly to when analyzing T1w images, the method was able to differentiate between patients and controls in almost all subregions, except for the posterior. This may be explained by the 5$mm$ spacing of the FLAIR images since it makes many images lack the mammilary bodies, or limit it to just one slice of the image. For this reason, the small AUROC values in this subregion are expected. Finally, we plotted the correlation among T1w and FLAIR normalized volumes (Figure~\ref{fig:exp4_corr}) to investigate whether H-SynEx exhibits consistency among them. The anterior subregion displays a moderate correlation (r=0.40 and r=0.50, respectively), and tuberal subregions have strong correlations (r=0.79 and r=0.80, respectively), both for controls and AD subjects. As expected, the posterior correlation is weak in both cases (r=0.11 and r=0.22). These results support the hypothesis that the method can be used in challenging resolutions and still detect differences among groups.

Although H-SynEx leverages randomized synthetic images to mitigate training bias, a limitation remains. The model's accuracy on unseen data can still be affected by the image contrast itself. For instance, when analysing Experiment 4.2, in both IXI and FSM there is only one label per subject, done on T1w images. Therefore the manual segmentations used to generate the quantitative results were not influenced by different contrasts, which may influence the final results. Also, we could demonstrate that the smallest subregions (anterior and posterior) had the biggest variability, especially in DC, an overlap measure known for being sensitive to small structures~\cite{taha2015metrics}.

To the best of our knowledge, we have presented the first automated method for hypothalamic subregion segmentation capable of working across different \textit{in vivo} MRI sequences and resolutions without retraining. By producing reliable and consistent segmentations, H-SynEx facilitates the analysis of the hypothalamus in various pre-existing datasets, whether in research or clinical settings. Our tool is publicly available and has the potential to increase our understanding of the roles played by the hypothalamus and its subregions in neurodegenerative diseases and other related conditions.

\section{Acknowledgements}
L.Rodrigues acknowledges the Coordination for the Improvement of Higher Education Personnel (88887.716540/2022-00). M. Bocchetta is supported by a Fellowship award from the Alzheimer’s Society, UK (AS-JF-19a-004-517). J.E.Iglesias acknowledges NIH 1RF1MH123195, 1R01AG070988, and a grant from the Jack Satter foundation. L. Rittner acknowledges CNPq 313598/2020-7 and FAPESP 2013/07559-3. S.Appenzeller acknowledges CAPES Print, CAPES 001 e BRAINN.

 \bibliographystyle{elsarticle-num} 
 \bibliography{cas-refs}





% \end{thebibliography}
\end{document}